# Optimization strategies for parallel CPU and GPU implementations of a meshfree particle method

Jose M. Domínguez, Alejandro J.C. Crespo* and Moncho Gómez-Gesteira

EPHYSLAB Environmental Physics Laboratory, Universidade de Vigo, Campus As Lagoas s/n, 32004, Ourense, Spain

E-mail addresses: jmdominguez@uvigo.es (J.M. Dominguez), alexbexe@uvigo.es (A.J.C. Crespo), mggesteira@uvigo.es (M. Gómez-Gesteira)

* Corresponding author: A.J.C. Crespo, alexbexe@uvigo.es, +34 988387255

**Abstract:** Much of the current focus in high performance computing (HPC) for computational fluid dynamics (CFD) deals with grid based methods. However, parallel implementations for new meshfree particle methods such as Smoothed Particle Hydrodynamics (SPH) are less studied. In this work, we present optimizations for both central processing unit (CPU) and graphics processing unit (GPU) of a SPH method. These optimization strategies can be further applied to many other meshfree methods. The obtained performance for each architecture and a comparison between the most efficient implementations for CPU and GPU are shown.

1. Introduction

Numerical methods are demanded as useful tools in engineering and science to solve complex problems. The main advantage of this approach is the capability to simulate complex scenarios without building costly scale models, and provide physical data that can be difficult, or even impossible, to measure in a real model. The physical governing equations of a numerical method can be solved with the help of a grid using an Eulerian description or can be solved without it following a Lagrangian description. We will here focus on meshfree methods, which make easier the simulation of problems with large deformations, complex geometries, nonlinear material behavior, discontinuities and singularities. In particular, the so called Smoothed Particle Hydrodynamics (SPH) method will be employed here as a benchmark. SPH was developed during the seventies in the field of astrophysics [1] and has been applied to different engineering fields such as fluid dynamics. There, problems involving free-surface flows, gravity currents, breaking waves and wave impact on structures are some of the most relevant applications.

Two important aspects must be considered in a numerical method. The first one is the correct implementation of the physical governing equations and the accuracy of the mathematical algorithms. The second one is directly related to the nature of hardware needed to execute the model. Each kind of platform (desktops, workstations, clusters…) used to perform numerical simulations present its own advantages and limitations. Parallelization methods and optimization techniques are essential to perform simulations at a reasonable execution time. Therefore, in order to obtain the best performance, the code must be optimized and parallelized as much as possible according to the available hardware resources.

Two different hardware technologies can be employed; Central Processing Units (CPUs) and Graphics Processing Units (GPUs). Current CPUs have multiple processing cores, making possible the distribution of the workload of a program among the different cores dividing the execution time. In addition, CPUs also present SIMD (Single Instruction, Multiple Data) which allows performing an operation on multiple data simultaneously. The parallelization task on CPUs can be performed by using MPI (Message Passing Interface) or OpenMP (Open Multi-Processing) as shown by [2] for large SPH simulations. In addition, parallel computing with multiple CPUs has been also used for SPH simulations in recent works (e.g. [3-6]). More recently, [7] presented a parallel SPH implementation on multi-core CPUs. On the other hand, research can be also conducted with the promising GPU technology for problems that previously required high performance computing (HPC). Initially, GPUs were developed exclusively for graphical purposes. However, the demands of the games and multimedia market forced a fast increase in performance that has led to parallel processors with a computing power in floating-point much higher than the CPU ones. Moreover, the emergence of languages such as the Compute Unified Device Architecture (CUDA) facilitates the programming of GPUs for general purpose applications. Therefore, recently the GPGPU programming (General Purpose on Graphics Processing Units) has experienced a strong growth in all fields of scientific computing. In the particular case of SPH, the first implementations of the method on GPU were carried out by [8] and [9]. Following the release of CUDA, different SPH models have been implemented on GPUs during the last few years [10,11].

SPHysics is a Smoothed Particle Hydrodynamics code mainly developed to deal with free-surface flow phenomena. It has been jointly developed by the Johns Hopkins University (US), the University of Vigo (Spain) and the University of Manchester (UK). SPHysics code can simulate complex fluid dynamics, including wave breaking, dam breaks, solid objects sliding into the water, wave impact on structures, etc. The first serial code was developed in FORTRAN (see [12] for a complete description of the code) showing its reliability and robustness for 2D [13-15] and 3D [16-18] problems. However, the main drawback of SPH models in general and of SPHysics in particular is their high computational cost. Implementations that exploit the parallelism capabilities of the current hardware should be developed to carry out the simulation of realistic domains at a reasonable runtime. A new code named DualSPHysics has been developed inspired by the formulation of the former FORTRAN SPHysics. The code, which can be freely downloaded from www.dual.sphysics.org, was implemented using both C++ and CUDA programming languages and can be executed both on CPUs and on GPUs.

It has been demonstrated that DualSPHysics achieves different speedups with different CUDA-enabled GPUs as shown in [11]. However, getting the most out of these parallel architectures is not trivial. This manuscript describes different strategies for CPU and GPU optimizations applied to DualSPHysics. A comparison of the performance obtained with CPU-multicore and GPU is also shown. This comparison is as fair as possible, since the most efficient implementations are adopted for each architecture. Furthermore, [19] remarks the importance of optimizations which are independent on the parallelization. Therefore, different improvements will be shown before and after-parallelization. In addition, these optimization techniques may be adopted by many other meshfree numerical methods. The reader is invited to use some of these optimization techniques and apply them to their own codes. Some optimizations are intrinsic to the SPH method but most of them are based on the Lagrangian nature of the meshfree methods. The GPU optimizations applied here to SPH present some of the well-known suggested basic optimizations described in the CUDA manuals [20], such as expose as much parallelism as possible, minimize CPU-GPU data transfers, optimize memory usage for maximum bandwidth, minimize code divergence, optimize memory access patterns, avoid non-coalesced accesses and maximize occupancy to hide memory latency.

## 2. SPH implementation

Smoothed Particle Hydrodynamics is a meshfree, Lagrangian, particle method for modeling fluid flows. Since the fluid is treated as a set of particles, the hydrodynamic equations of motion are integrated on each particle. The conservation laws of continuum fluid dynamics written in the form of partial differential equations need to be transformed into the Lagrangian formalism (particles). Using an interpolation function that provides the weighted estimate of the field variables at a discrete point (particle), the integral equations are evaluated as sums over neighboring particles. Only particles located at a distance shorter than $h$, the so called smoothing length, will interact. Thus, physical magnitudes (position, velocity, mass, density, pressure) are computed for each particle as an interpolation of the values of the nearest neighboring particles. The governing equations expressing conservation laws of continuum fluid dynamics are computed for pair-wise interactions of fluid particles, however the interaction with boundary particles needs different computations. Therefore, two different types of particles are considered; fluid particles and boundaries.

A more complete description of the SPH method can be found at [21-23] and the particular formulation of DualSPHysics code is presented in [11].

In terms of implementing the code, the DualSPHysics code is organized in three main stages that are repeated each time step: (i) creating a neighbor list; (ii) computing particle interactions for momentum and continuity conservation equations; and (iii) integrating in time to update all the physical properties of the particles in the system (SU). A diagram of the SPH code can be seen in Fig. 1. The first stage is creating the neighbor list (NL). As we mentioned above, particles only interact with neighboring particles located at a distance less than *h*. Thus, the domain is divided into cells of size (*h*×*h*×*h*) to reduce the neighbor search to only the adjacent cells and the cell itself. The Cell-linked list described in [24] was implemented in DualSPHysics. In fact, two different lists were created; the first one with fluid particles and the other one with boundary particles. The second stage is the particle-particle interaction (PI), where each particle checks which particles contained in the adjacent cells and in its own cell are real neighbors (placed at a distance shorter than *h*). Then, the equations of conservation of mass and momentum are computed for the pair-wise interaction of particles. The interaction boundary-boundary is not necessary. Thus, only fluid-fluid (F-F), fluid-boundary (F-B) and boundary-fluid (B-F) interactions should be carried out. Finally, the system is updated (SU) so, once the interaction forces between particles are calculated, the physical quantities are integrated in time at the next time step.

The testbed used in this work is the gravity induced collapse of a water column (see Fig. 2). Similar cases were simulated to validate the SPHysics [16] and the DualSPHysics [11] codes. This testcase will be used to analyze the performance of the optimizations and implementations presented in the following sections.

As mentioned above, the SPH method is expensive in terms of computational time. For example, a simulation of the dam break evolution during 1.5s of physical time using 300,000 particles takes more than 15 hours on a single-core machine (before the optimizations described in this paper). The first limitation is the small time step ($10^{-6}$-$10^{-5}$ s) imposed by forces and velocities [25]. Thus, in this case, more than 16,000 steps are needed to complete the 1.5s of physical time. On the other hand, each particle interacts with more than 250 neighbors, which implies a large number of interactions (operations) in comparison with the methods based on a mesh (Eulerian methods). In this case, the particle interaction takes 99% of the total computational time when executed on a single-core CPU and more than 95% on a GPU. Thus, all the efforts to increase the performance of the code must be focused on reducing the execution time of the particle interaction stage.

Finally, we should mention that SPHysics [12, 23] allows numerous parameterizations to compute boundary conditions, viscosity, time stepping… In this particular study, the chosen options are summarized in Table I.

3. **Strategies for CPU optimization**

The key details about the CPU implementation should be addressed before describing the main optimizations on CPU of the code. The physical quantities of each particle (position, velocity, density…) are stored in arrays. During the NL stage, the cell to which each particle belongs is determined. This makes possible to reorder the particles (and the arrays with particle data) following the order of the cells. Thus, if particle data are closer in the memory space, the access pattern is more regular and more efficient. Another advantage is the ease to identify the particles that belongs to a cell by using a range since the first particle of each cell is known. In this way, the interaction between particles is carried out in terms of the interaction between cells. All the particles inside a cell interact with all the particles located in the same cell and in adjacent cells. Force computations between two particles will be carried when they are closer than the interaction range (*h*).

The optimizations to be applied to the CPU implementation of DualSPHysics are:

3.1 Applying symmetry to particle-particle interaction (A).

When the force, $f_{ab}$, exerted by a particle, *a*, on a neighbor particle, *b*, is computed, the force exerted by the neighboring particle on the first one can be known since it has the same magnitude but opposite direction ($f_{ba} = -f_{ab}$). Thus, the number of interactions to be evaluated can be reduced by two, which decreases the computational time. For this purpose, in 3D, each cell only interacts with 13 cells and, partially, with itself (symmetry is also applied for the particles inside the same cell), instead of 27 as shown in Fig. 3.

3.2 Splitting the domain into smaller cells (B).

Usually, in particle methods, the domain is split into cells of size ($h \times h \times h$) to reduce the neighbor search to only the adjacent cells. Thus, in 3D and without considering symmetry, a volume of $27h^3$ is searched for every cell to look for potential neighbors. This volume is considerably higher than the volume of the sphere of radius *h* around the target particle, *a,* where its real neighbors are placed ($V_{sphere}=(4/3)\pi h^3 \sim 4.2h^3$). This can be generalized to any division of the computational domain into cells of side *h/n*. Thus the ratio between the searched volume and the sphere volume becomes $(2+(1/n))^3/((4/3)\pi)$, which tends asymptotically to $6/\pi$ when *n* goes to infinity. Thus, a suitable technique to diminish the number of *false* neighbors would be to reduce the volume of the cell. Unfortunately, each cell requires the storage of information to identify its beginning, end and number of particles, which prevent the use of large *n* values. A balance between decreasing the searching volume and limiting memory requirements should be found. According to our experience, *n* values on the order of 2 are recommended. Figure 4 shows the comparison between dividing the domain into cells of side *h* (*n*=1) and side *h/2* (*n*=2).

3.3 Using SSE instructions (C).

The current CPUs have special instruction sets (SSE, SSE2, SEE3…) of SIMD type (Single Instruction, Multiple Data) that allow performing operations on data sets. A basic operation (addition, subtraction, multiplication, division, comparison…) of four real numbers (in single precision) can be executed simultaneously. Another advantage is the straightforward translation to machine-code providing a higher performance. However this optimization also presents two disadvantages: first, coding is quite cumbersome and, second, the technique can only be applied to specific cases where the calculations are performed in packs of 4 values. Although modern compilers implement the automatic use of these SIMD instructions, [11] emphasize the need of making an explicit vectorization of the computations to obtain the best performance on the CPU. Therefore, these instructions are applied to the interaction between particles that were previously grouped into packs of 4 to compute forces simultaneously. An example of a simplified pseudocode can be seen in Fig. 5.

3.4 Multi-core programming with OpenMP (D).

Current CPUs have several cores or processing units, so it is essential to distribute the computation load among them to maximize the CPU performance and to accelerate the SPH code. As mentioned above, there are two main options to implement a parallel code in CPU, namely MPI and OpenMP. MPI is particularly suitable to distribute memory systems where each processing unit has access only to a portion of the system memory and processes need to exchange data by passing messages. However the architecture used in this work consists on a shared memory system, where each process can directly access to all memory without the extra cost of the message passing in MPI. OpenMP is a portable and flexible programming interface whose implementation does not involve major changes in the code. Using OpenMP, multiple threads for a process can be easily created. These threads are distributed among all the cores of the CPU sharing the memory. So, there is no need to duplicate data or to transfer information between threads. Due to these reasons, OpenMP was used in DualSPHysics.
Several parts of the SPH code can be parallelized, which is especially important for force calculation, since it is the most expensive part of the code. The minimum execution unit of each thread is the cell, so that all particles of the same cell are processed sequentially. Neighboring particles are searched in the surrounding cells and the particle interaction is computed. Two basic problems appear in this parallel programming; first, the concurrent access to the same memory positions for read-write particle forces, which gives rise to unexpected results, and, second, the load balancing to distribute equally the work among threads. Three different approaches were proposed to avoid concurrent accesses and obtain load balancing:

*Asymmetric*: Concurrent access occurs in force computation when applying symmetry, since the thread that computes the summation of the forces on a given particle is also computing the forces on the particles placed in the neighborhood of the first one. Nevertheless, these neighboring particles may be simultaneously processed by another thread. To avoid this conflict, symmetry is not applied in a first approach. The load balancing is achieved by using the dynamic scheduler of OpenMP. Cells can be assigned (in blocks of 10) to the threads as they run out of workload. Fig. 6 shows an example of dynamic distribution of cells (in blocks of 4) among 3 execution threads according to the execution time of each cell, which depends on the number of neighboring particles. The main advantage is the ease of implementation, being the main drawback the loss of symmetry.

*Symmetric*: In this approach, the dynamic scheduler of OpenMP is also employed distributing cells in blocks of 10 among different threads. The difference with the previous case lies in the use of the symmetry in the computation of the particle-particle interaction. Now the concurrent memory access is avoided since each thread has its own memory space, where the forces on each particle are accumulated. Thus, the final value of the interaction force of each particle is obtained by combining the results once all threads have finished. This final value is also computed by using multiple threads. The advantage of this approach is the use of the symmetry in all the interactions and the easy implementation of the load dynamic balancing. The main drawback is the increase in memory requirements, which depends on the number of threads. For example, memory increases by a factor of 2 when passing from 1 to 8 threads in our testcase.

*Slices*: The domain is split into slices. Symmetry is applied to the interactions among cells that belong to the same slice, but not to the interactions with cells from other slices. Thus, symmetry is used in most of the interactions (depending on the width of the slices). The thickness of the slices is adjusted to distribute the runtime of the particle interactions within each slice (dynamic load balancing). The division is periodically updated to keep the slices as balanced as possible. This thickness is adjusted according to the computation time required for each slice during the last time steps, which allows a more correct dynamic load balancing. The main drawbacks are the higher complexity of the code and the higher runtime associated to the dynamic load balancing.

## 4. Strategies for GPU optimization

As mentioned in Section 2, the execution of particle interactions for the present testcase takes 99% of the total runtime in a single-core CPU. For the GPU implementation, we will start from a hybrid implementation where only the PI stage is implemented on the GPU (partial GPU implementation). The memory of the GPU device is physically independent of the CPU memory so, when data are processed on the GPU, two transfers (CPU to GPU and GPU to CPU) are needed every time step. Thus, different stages must be carried out to compute one time step: (i) generating the neighbor list on CPU, (ii) transferring particle data and cell information to the GPU, (iii) computing forces on GPU, (iv) recovering the forces calculated on GPU and (v) updating magnitudes of the particles at the next time step (see left panel in Fig. 7).

The implementation of the particle interactions on GPUs is different from the CPU one. Instead of computing the interaction of all particles of one cell with the neighboring cells as implemented on CPUs, now each particle looks for all its neighbors sweeping all the adjacent cells. Thus, particle interactions can be implemented on the GPU for only one particle using one execution CUDA thread to compute the force resulting from the interaction with all its neighbors. Using this approach, the symmetry of the force computation cannot be applied since several threads could be trying to modify simultaneously the same memory position giving rise to an error. This problem could be avoided by using synchronization barriers, nevertheless it would be no longer efficient due to its high computational cost.

This implementation presents different problems to be solved:
- *CPU-GPU transfers*: Since this is a partial implementation on GPU, memory communications must be established between CPU and GPU at each time step. Therefore, the performance of the code is significantly reduced.
- *Code divergence*: GPU threads are grouped into sets of 32 named *warps* in CUDA language. When a task is being executed over a warp, the 32 threads carry out this task simultaneously. However, due to conditional flow instructions in the code, not all the threads will perform the same operation, so the different tasks are executed sequentially, giving rise to a significant loss of efficiency. This divergence problem appears during particle interaction since each thread has to evaluate which potential neighbors are real neighbors before computing the force.
- *No coalescent memory accesses*: The global memory of the GPU is accessed in blocks of 32, 64 or 128 bytes, so the number of accesses to satisfy a warp depends on how grouped data are. In particle interaction, although particle data are reordered according to the cells they belong to, a regular memory access is not possible since each particle has different neighbors and therefore each thread will access to different memory positions which may, eventually, be far from the rest of the positions in the warp.
- *No balanced workload*: Warps are executed in *blocks* in the CUDA terminology. When a block is going to be executed, some resources are assigned and they will not be available for other blocks till the end of the execution. So, since each thread may have a different number of neighbors, a thread may need to perform more interactions than the rest. Thus, the warp can be under execution while the rest of threads of the same warp, or even of the block, can have finished. Thus, the performance is reduced due to inefficient use of the GPU resource.

In order to avoid or minimize the problems previously described, the following optimizations were developed:

4.1 Full implementation on GPU (A).

To minimize the CPU-GPU transfers, all the SPH method can be implemented on GPU keeping data in the GPU memory (see right panel of Fig. 7). Therefore, the CPU-GPU communications are drastically reduced and only some particular results will be recovered from GPU at some time steps. Moreover, when the other two main stages of the SPH method (neighbor list and system update) are implemented on GPU, the computational time devoted to these processes decreases.

As previously described for CPUs, particle data (position, velocity…) are stored in arrays but, in this case, the arrays are stored on the GPU memory. During the neighbor list stage on GPU, the cells where the particles are located are calculated first. Then, the implementation of the *radixsort* algorithm from NVIDIA [31] is used to obtain the position of the particles when they are reordered according to cells, so all the arrays with particle data are reordered. Finally, a new array is created to determine which particles belong to a given cell. Thus, the array *CellBeginEnd* contains the first and the last particle of every cell. Fig. 8 shows an example of how this array is created starting from the particle distribution depicted in the left panel of the figure. The method to create *CellBeginEnd* is described in [32], which has proven to be efficient even for empty cells.

The system update can be easily parallelized on a GPU since only the physical properties of the particles must be calculated at the next time step. The major arduousness is computing the maximum and minimum values of some variables (force, velocity and sound speed) needed to estimate the value of the variable time step [25]. This calculation is optimized using a reduction algorithm for GPU based on [33]. This algorithm allows obtaining the maximum or minimum values of a huge data set taking advantage of the parallel programming in GPUs.

4.2 Maximizing the occupancy of GPU (B).

Occupancy is the ratio of active warps to the maximum number of warps supported on a multiprocessor of the GPU or Streaming Multiprocessor (SM). Since the access to the GPU global memory is irregular during the particle interaction, it is essential to have the largest number of active warps in order to hide the latencies of memory access and maintain the hardware as busy as possible. The number of active warps depends on the registers required for the CUDA kernel, the GPU specifications (see Table 2) and the number of threads per block.

Using this optimization, the size of the block is adjusted according to the registers of the kernel and the hardware specifications to get the maximum occupancy. Fig. 9 shows the obtained occupancy for different number of registers and for different compute capabilities of the GPU card when using 256 threads and using other block sizes. For example, the occupancy of a GPU sm13 (compilation with compute capability 1.3) for 35 registers is 25% (dashed blue line) using 256 threads, but it can be 44% (solid blue line) using 448 threads.

4.3 Reducing global memory accesses (C).

When computing the SPH forces during the PI stage, the six arrays described in Table 3 are used. The arrays *csound*, *prrhop* and *tensil* are previously calculated for each particle using *rhop* to avoid calculating them for each interaction of the particle with all its neighbors. The number of memory accesses in the interaction kernel can be reduced by grouping part of these arrays (*pos+press* and *vel+rhop* are combined to create two arrays of 16 bytes each one) and prevent reading values that can be calculated from other variables (*csound* and *tensil* are calculated from *press*). Thus, the number of accesses to the global memory of the GPU is reduced from 6 to 2 and the volume of data to be read from 40 to 32 bytes.

4.4 Simplifying the neighbor search (D).

During the GPU execution of the interaction kernel, each thread has to look for the neighbors of its particle sweeping through the particles that belong to its own cell and to the surrounding cells, a total of 27 cells since symmetry cannot be applied. However, this procedure can be optimized when simplifying the neighbor search. This process can be removed from the interaction kernel if the range of particles that could interact with the target particle is previously known. Since particles are reordered according to the cells and cells follow the order of X, Y and Z axis, the range of particles of three consecutive cells in the X-axis ($cell_{x,y,z}$, $cell_{x+1,y,z}$ y $cell_{x+2,y,z}$) is equal to the range from the first particle of $cell_{x,y,z}$ to the last of $cell_{x+2,y,z}$. Thus, the 27 cells can be defined as 9 ranges of particles. The 9 ranges are colored in Fig. 10.

The interaction kernel is significantly simplified (see pseudocode in Fig. 11), when these ranges are known in advance. Thus, the memory accesses decrease and the number of divergent warps is reduced. In addition, GPU occupancy increases since less registers are employed in the kernel. The main drawback is the higher memory requirements due to the extra 144 bytes needed per cell.

4.5 Adding a more specific CUDA kernel of interaction (E).

Initially, the same CUDA kernel was used to calculate all interaction forces B-F, F-B and F-F. However, symmetry in the force computation cannot be efficiently applied and the best option is implementing a specific kernel for the B-F interaction (boundary-fluid) because only a subset of the fluid particles is required to be computed for the boundaries. The effect of this optimization on the overall performance is negligible when the number of boundary particles is small in comparison with the number of the fluid ones.

On the other hand, the access to the global memory of the GPU is two orders of magnitude slower than the access to other registers. In order to minimize these accesses, each thread starts storing all its particle data in registers, so the thread only needs to read data corresponding to the neighbor particles. The same approach is applied to store the forces, which are accumulated in registers and written in global memory at the end. As described before, there are two types of particles (boundaries and fluids), so there are three interactions to calculate all the forces (F-F, F-B and B-F). Therefore, data of the fluid particles associated to the threads are read twice (when fluid particles interact with other fluid particles and when they interact with boundaries) and the same occurs when writing results in the global memory. A way to avoid this problem is carrying out the interaction F-F and F-B in the same CUDA kernel with only one initial data load and one final writing of the results instead of two.

4.6 Division of the domain into smaller cells (F).

As described in the optimization applied in the CPU implementation, the procedure consists in dividing the domain into cells of size *h/2* instead of size *h* in order to increase the percentage of real neighbors. Using cells of size *h* on the GPU implementation, the number of pair-wise interactions decreases. The disadvantage is the increase in memory requirements since the number of cells is 8 times higher and the number of ranges of particles to be evaluated in the neighbor search increases from 9 to 25 (using 400 bytes per cell). Fig. 12 represents the memory needed to allocate different number of particles. This amount of memory increases drastically when decreasing the size of the cells.

## 5  Results

The DualSPHysics code will be used to run the testcase described above (see Fig. 2). In this section, the results of applying all the optimizations developed in this work are presented. Thus, the relative performance of the CPU and GPU implementations at different levels of optimization is studied.

The system used for the CPU performance testing:
- *Hardware*: Intel® Core ™ i7 940 at 2.93 GHz (4 physical cores, 8 logical cores with Hyper-threading, with 6 GB of 1333 MHz DDR3 RAM)
- *Operating system*: Ubuntu 10.10 64-bit
- *Compiler*: GCC 4.4.5 (compiling with the option –O3)

The system used for the GPU performance testing:
- *Hardware1*: NVIDA GTX 480 (15 Multiprocessors , 480 cores at 1.37 GHz with 1.5 GB of 1848 MHz GDDR5 RAM and compute capability 2.0)
- *Hardware2*: NVIDA Tesla 1060  (30 Multiprocessors, 240 cores at 1.3 GHz with 4 GB of 1600 MHz GDDR3 RAM and compute capability 1.3)
- *Operating system*: Debian GNU/Linux 5.0 (Lenny) 64-bit.
- *Compiler*: CUDA 3.2 (compiling with the option –use_fast_math).

Both CPU and GPU results show the speedup obtained when comparing the performance of the optimized version with the basic version (without any of the described optimizations). The performance is measured as the number of time steps computed per second.

Fig. 13 shows the achieved speedup on CPU for different number of particles (N) when applying the three first optimization strategies explained in Section 3; symmetry in particle interaction (A), division of the domain into cells of size *h/2* (B) and use of SSE instructions (C). The blue line in Fig. 13 shows the speedup obtained using symmetry and the red line includes the speedup when using SSE instructions and symmetry, the value in parentheses is the cell size. Using 300,000 particles, a maximum speedup of 2.3x is obtained using these CPU optimizations when compared to the version of the code without optimizations.

The speedup obtained with the multi-core implementation on CPU of the SPH code (optimization D in Section 3) for different number of particles is observed in Fig. 14. In the figure, the performance of the different OpenMP implementations (using 8 threads) is compared with the most efficient single-core version (that includes symmetry, SSE instructions and cell size equal to *h/2*). The most (less) efficient implementation is *Symmetric* (*Asymmetric*). A speedup of 4.5x is obtained with *Symmetric* when using 8 threads. Note, that the optimal number of threads with this hardware is 8 since the code is executed on a CPU activating Hyper-threading that has 8 logical cores over 4 physical cores. The approaches that divide the domain into slices (*Slices*) offer a higher performance when increasing the number of particles since the number of cells also increases, which allows a better distribution of the workload among the 8 execution threads. Using *Slices*, the direction of fluid movement is not important. Similar performance is achieved when creating the slices in X or Y-direction, since the workload is distributed equally among the slices.

Fig. 15 shows the achieved speedup on CPU using all the proposed optimizations in comparison to the basic single-core version without optimizations. Thus, the speedups with the most efficient version of OpenMP (*Symmetric*) and the single-core code (including all optimizations) are shown in Fig. 15. For a simulation involving 300,000 particles, the optimized single-core implementation outperforms the basic single-core by 2.3x while the optimized OpenMP implementation using 8 logical threads is 10.3 times faster. For simulations of systems larger than 150,000 particles, using Hyper-threading speeds up the code in about 20%.

Now, the different GPU results will be analyzed. Fig. 16 (Fig. 17) shows the speedup achieved on the GPU card GTX 480 (Tesla 1060) in comparison to the basic GPU version without the implementation of the different optimization strategies described in Section 4. Note that each line of the figures represents the speedup of each optimization including all previous ones. Thus, the line that corresponds to the optimization E (implementing a more specific kernel according to the type of interaction) includes the previous optimizations; full implementation on GPU (A), maximizing the occupancy (B), reducing global memory accesses (C) and simplifing the neighbor search (D). The speedup obtained when using optimizations E on the GTX 480 is 2.1x simulating for a run involving 2 million particles. Using the optimization F (division of the domain in cells of size *h/2*), the maximum number of particles allowed in a GTX 480 is only 1.8 million since its capacity is only 1.5 GB. Thus, for 1.8 million particles, the fully optimized GPU code for the GTX 480 is 2.6 times faster than the basic GPU version without optimizations. In the case of Tesla 1060, whose capacity is 4 GB, the achieved speedup was 2.4.

One of the most important optimizations is the full implementation of the SPH code on GPU. When neighbor list (NL), particle interaction (PI) and system update (SU) are implemented on GPU, the CPU-GPU data transfer is avoided in each time step. Figure 18 shows the computational runtimes using the GTX 480 for different GPU optimizations (partial, full and optimized) simulating 500,000 particles of the testcase. *Partial GPU* implementation corresponds to the initial version where only the PI stage is implemented on GPU, *full GPU* is the first version where the three stages of the SPH code are executed on GPU and *optimized GPU* is the final version including all the proposed optimizations described in Section 4. It can be observed that the time dedicated to the CPU-GPU data transfer in the partial implementation is 9.4% of the total runtime. Once the SPH code is totally implemented on GPU, the CPU-GPU communications are not necessary at each time step. The runtimes of the NL and SU stages decrease when these two parts of the code are also implemented on GPU. Finally, after applying all the developed optimization strategies, the computational time of the PI stage is reduced in about 40%.

The comparison between CPU and GPU can be observed in Fig. 19. The figure shows the speedup achieved with the most efficient implementation on GPU versus the optimized multi-core implementation on CPU (*Symmetric* with 8 threads) and the optimized single-core implementation. The speedups are shown for GTX 480 (solid lines) and Tesla 1060 (dashed lines). It is observed that the speedups with GTX 480 (Fermi technology) are twice those obtained with Tesla 1060, which belongs to a previous generation of GPU cards.

Table 4 summarizes the execution runtimes, the number of computed steps and the achieved speedups for the testcase described above. Note that the speedup is the ratio between the numbers of time steps computed per second by the different versions. Thus, for example, for one million particles, the performance of the CPU is 0.2 time steps per second using the single-core version and 0.8 using the multi-core version, while 10.1 time steps per second can be computed with a GPU GTX 480. The whole simulation takes one day, 16 hours and 45 min on the Intel® Core ™ i7 and only 42 min on the GTX 480, resulting in a speedup of 56.2x.

As explained before, the fastest GPU implementation also presents the highest memory requirements, therefore the maximum number of particles that can be simulated in a GTX 480 is only 1.8 million. Therefore, three different versions of the DualSPHysics code are implemented to avoid this limitation: the first one contains all the GPU optimizations (A+B+C+D+E+F) and it is named *FastCells(h/2)*, the second one, named *SlowCells(h/2)*, is implemented without the optimization D (simplify the neighbor search) that it is name and the third version, named *SlowCells(h)*, is implemented without the optimizations D and F (division of the domain in cells of size *h/2*). The memory usage for these three different GPU versions can be seen in Fig. 20. Note the black solid line represents the limit of memory that can be allocated on a GTX 480 (less than 1.4 GB). Using all these different versions that will be applied automatically during the execution, depending on the memory requirements, DualSPHysics allows simulating up to 9 million particles with a GTX 480 and more than 25 million with a Tesla 1060.

On the other hand, the execution times corresponding to these three GPU versions (*FastCells(h/2)*, *SlowCells(h/2)* and *SlowCells(h)*) and the times of the single-core and multi-core CPU versions are summarized in Fig. 21.

In summary, these three different GPU versions are available in DualSPHysics and the one to be executed is automatically selected by the code, unless indicated otherwise, depending on the memory requirements of the simulation.

# 6 Conclusions and future work

A code based on the SPH method has been developed to deal with free-surface flow problems. SPH is a particle meshless method with the benefits and problems inherent to its Lagrangian nature. A CPU-GPU solver named DualSPHysics is used to simulate a dam break flow impacting on a structure. Different strategies for CPU and GPU optimizations have been developed to speed up the results. The main advantages and disadvantages of each optimization are addressed.

Four optimizations are developed for the CPU implementations. The first one applies symmetry in particle interactions, the second one divides the domain into smaller cells, the third one uses SSE instruction and the fourth one uses OpenMP to implement multi-core executions. Three different approaches of the multi-core implementations are presented. The most efficient version uses the dynamic scheduler of OpenMP to achieve the load dynamic balancing and applies symmetry to particle interaction. Thus, the optimized single-core version is 2.3 times faster than the basic version without optimizations and the most efficient OpenMP implementation outperforms the basic single-core by 10.3x using the available 8 logical cores provided by the CPU hardware used in this study.

Six optimizations are developed for the GPU implementations; full implementation of the code on GPU to expose as much parallelism as possible, maximization of occupancy to hide memory latency, reduction of global memory accesses to avoid non-coalesced memory accesses, simplification of the neighbor search, optimization of the interaction kernel and division of the domain into smaller cells to reduce code divergence. The optimized GPU version of the code outperforms the GPU implementation without optimizations by a factor on the order of 2.6x.

The GPU parallel computing developed here can accelerate serial SPH codes with a speedup of 56.2x when using the Fermi card. The achieved performance can be compared to the large cluster machines [6], which are expensive and hard to maintain.

This work tries to help not only the users and developers of SPH methods, but also to those working with other particle methods. Thus, the adaptation of these optimizations to other meshless methods should be straightforward for a programmer with some experience on C++ and CUDA.

**Acknowledgements**

This work was supported by Xunta de Galicia under project Programa de Consolidación e Estructuración de Unidades de Investigación Competitivas (Grupos de Referencia Competitiva) and also financed by European Regional Development Fund (FEDER). We want to acknowledge Orlando Garcia Feal and Diego Perez Montes for their technical help.

**Figures captions**

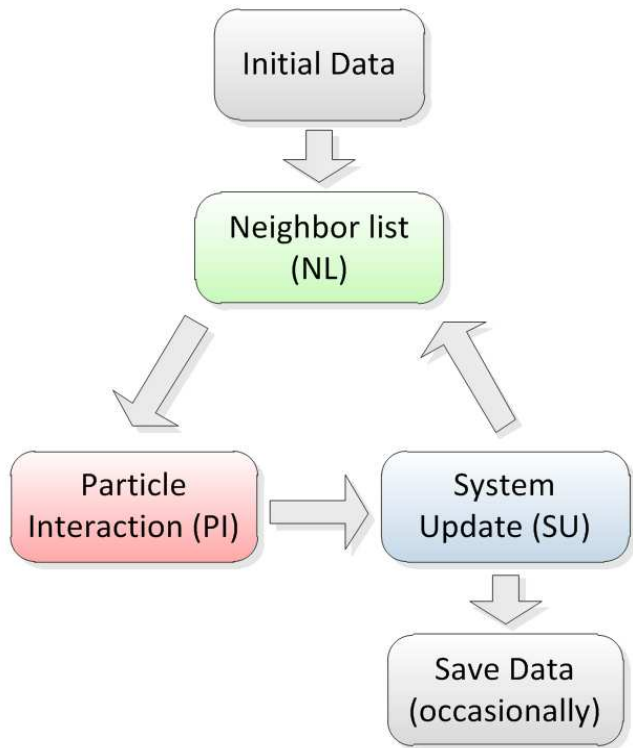

**Fig. 1**. Conceptual diagram summarizing the implementation of a SPH code.

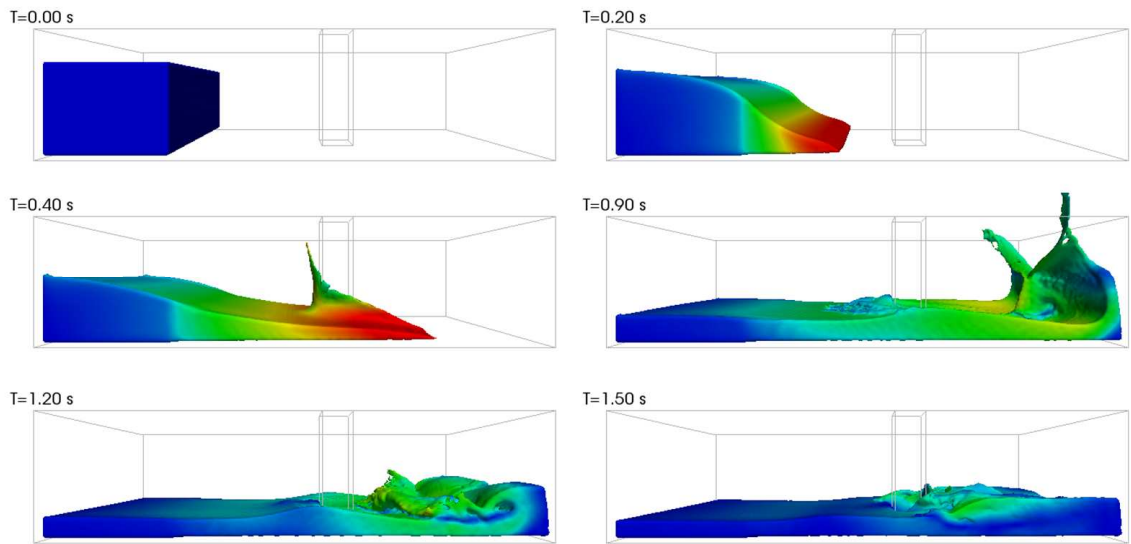

**Fig. 2**. Different instants of the dam break evolution.

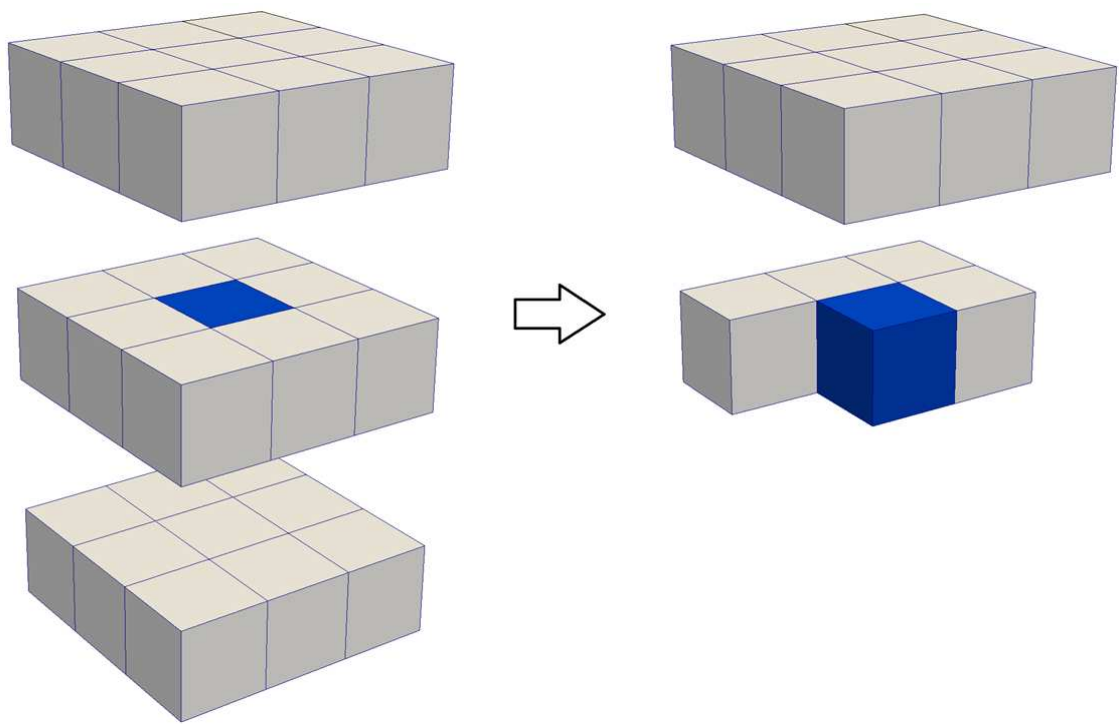

**Fig. 3**. Interaction cells in 3D without (left) and with (right) symmetry in particle interactions. Each cell interacts with 14 cells (right) instead of 27 (left).

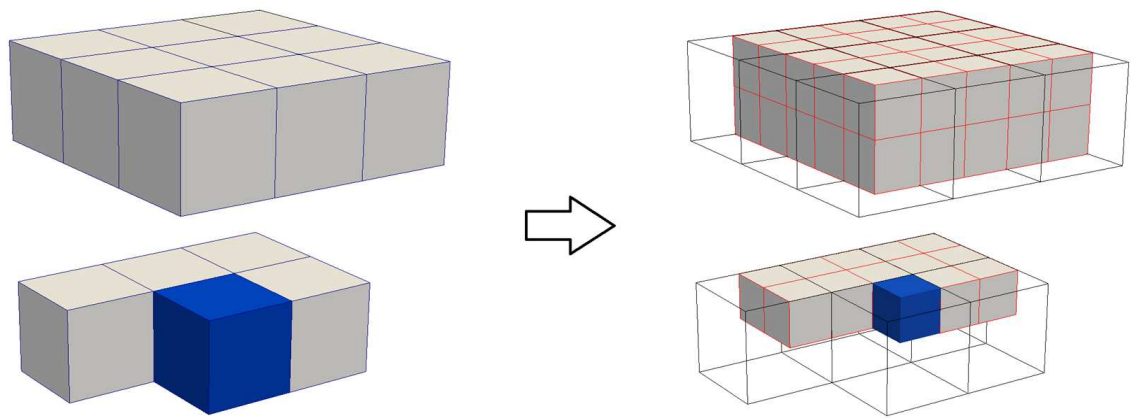

**Fig. 4**. Sketch of 3D interaction with close cells using symmetry. The volume searched using cells of side *h* (left panels) is bigger than using cells of side *h/2* (right panels).

```cpp
for(i=ibegin;i<iend;i++){
  for(j=jbegin;j<jend;j++){
    if(Distance between particle[i] and particle[j] < 2h)ComputeForces(i,j);
  }
}
```

```cpp
int npar=0;
int particlesi[4],particlesj[4];
for(int i=ibegin;i<iend;i++){
  for(int j=jbegin;j<jend;j++){
    if(Distance between particle[i] and particle[j] < 2h){
      particlesi[npar]=i; particlesj[npar]=j;
      npar++;
      if(npar==4){
        ComputeForcesSSE(particlesi,particlesj);
        npar=0;
      }
    }
  }
}
for(int p=0;p<npar;p++)ComputeForces(particlesi[p],particlesj[p]);
```

**Fig. 5**. Pseudocode in C++ showing the force computation between the particles of two cells without vectorial instructions (up) and grouping in blocks of 4 pair-wise of interaction using SSE instructions (down).

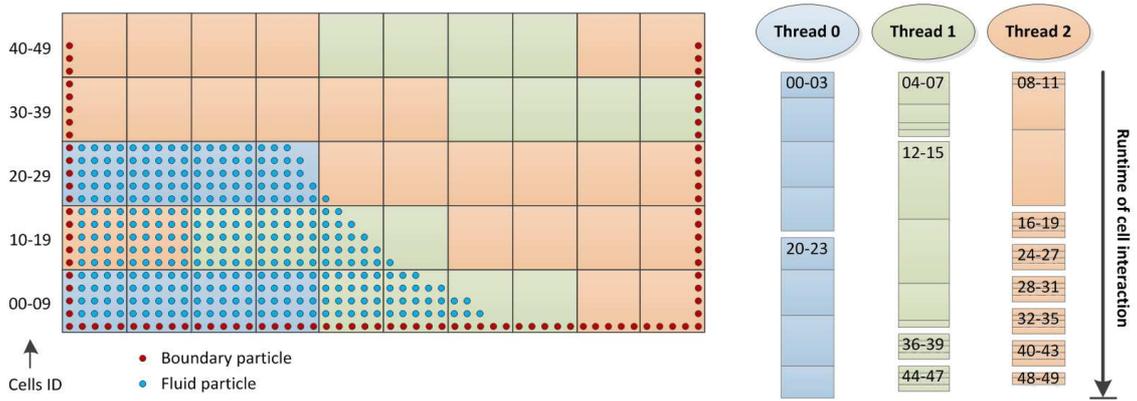

**Fig. 6**. Example of dynamic distribution of cells (in blocks of 4) among 3 execution threads according to the execution time of each cell.

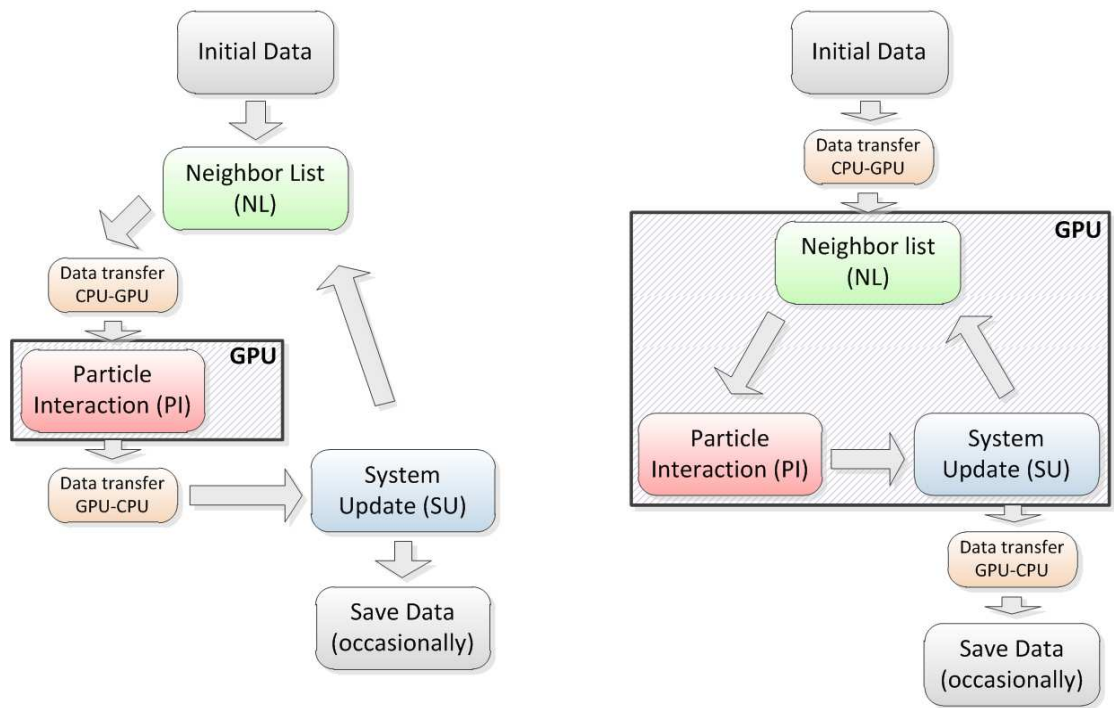

**Fig. 7**. Conceptual diagram of the partial (left) and full (right) GPU implementation of the SPH code.

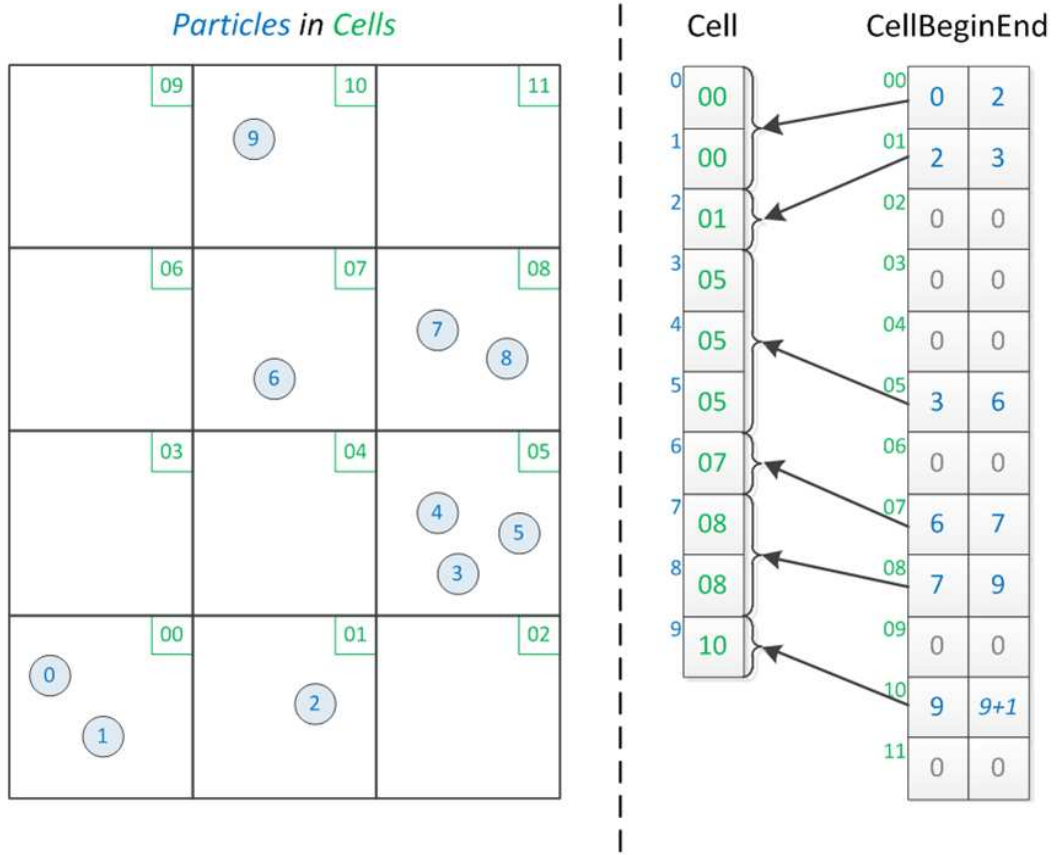

**Fig. 8**. Example of how *CellBeginEnd* is created starting from the particle distribution in cells depicted in the left panel.

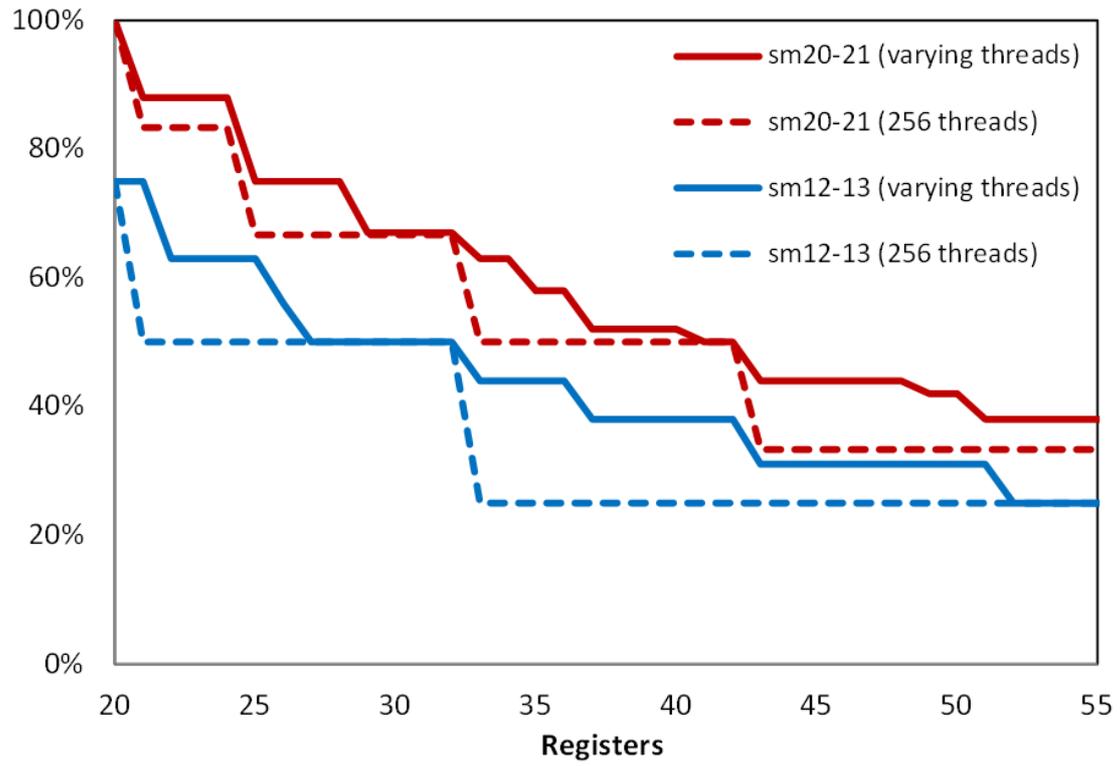

**Fig. 9**. Occupancy of the GPU for different number of registers with a variable and a fixed block size of 256 threads.

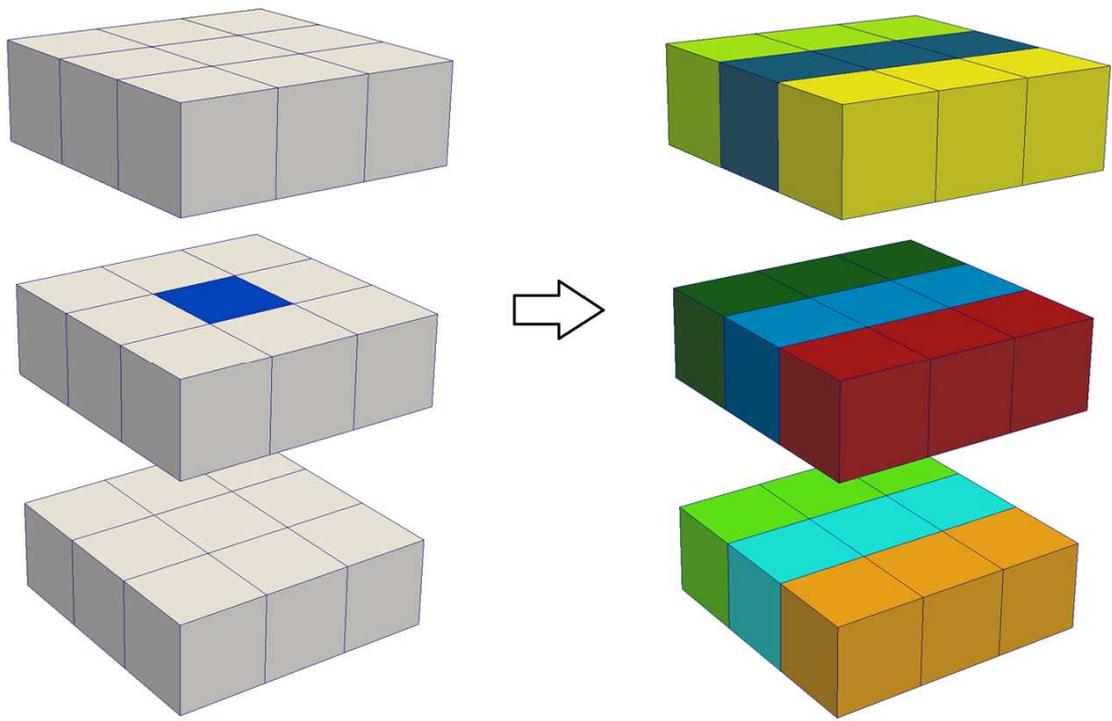

**Fig. 10**. Interaction cells in 3D without symmetry but using 9 ranges of three consecutive cells (right) instead of 27 cells (left).

```
For CUDA thread
  Load data of particle i (pos,vel,rhop,...) in registers
  Calculate minimum and maximum X coordinate for cells interaction (xmin and xmax)
  Calculate minimum and maximum Y coordinate for cells interaction (ymin and ymax)
  Calculate minimum and maximum Z coordinate for cells interaction (zmin and zmax)
  For cz=zmin to zmax
    For cy=ymin to ymax
      Calculate range of particles between Cell[xmin,cy,cz] and Cell[xmax,cy,cz]
      For each particle j in range
        If the distance between A and B is less than or equal to 2h
          Compute force between particle i and particle j
        End
      End
    End
  End
  Copy results stored in registers to global memory
End
```

```
For CUDA thread
  Load data of particle i (pos,vel,rhop,...) in registers
  For range=1 to 9 (there is 9 interaction ranges per cell)
    For each particle j in cellranges[cell i,range]
      If the distance between A and B is less than or equal to 2h
        Compute force between particle i and particle j
      End
    End
  End
  Copy results stored in registers to global memory
End
```

**Fig. 11**. Pseudocode in CUDA for the force computation in GPU between one particle and its neighbors using the array *CellBeginEnd* (up) and using the ranges of interaction previously calculated for each cell (down).

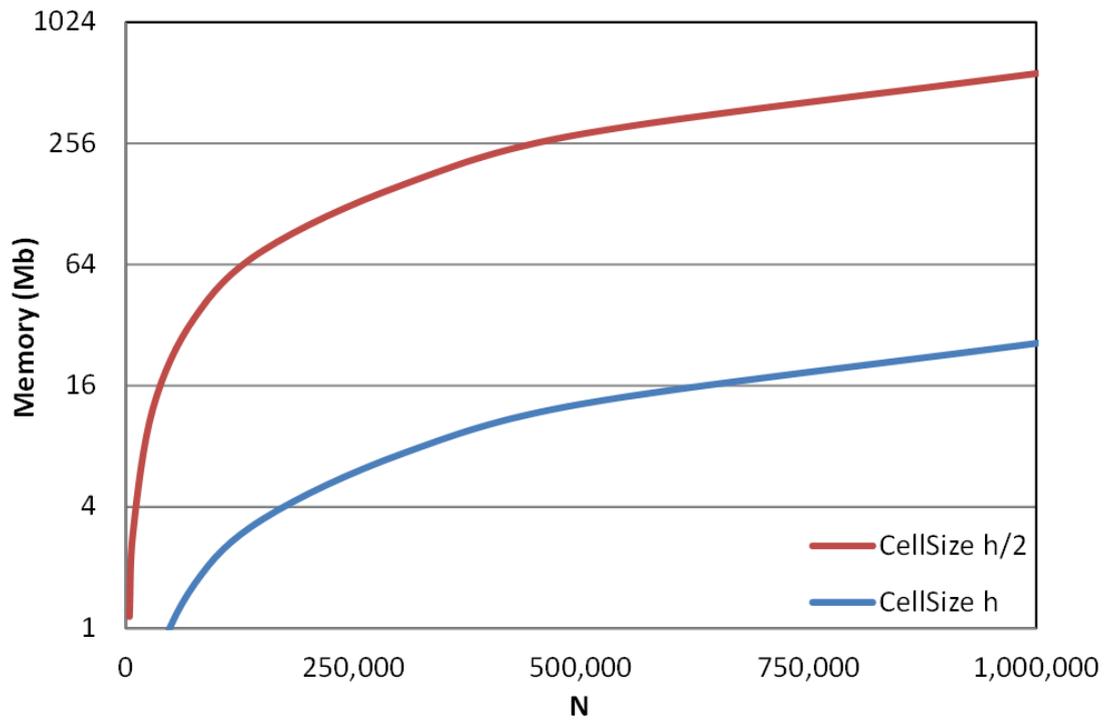

**Fig. 12**. Memory requirements to allocate the ranges of interaction using a cell size of *h/2* (red line) and *h* (blue line) for the testcase.

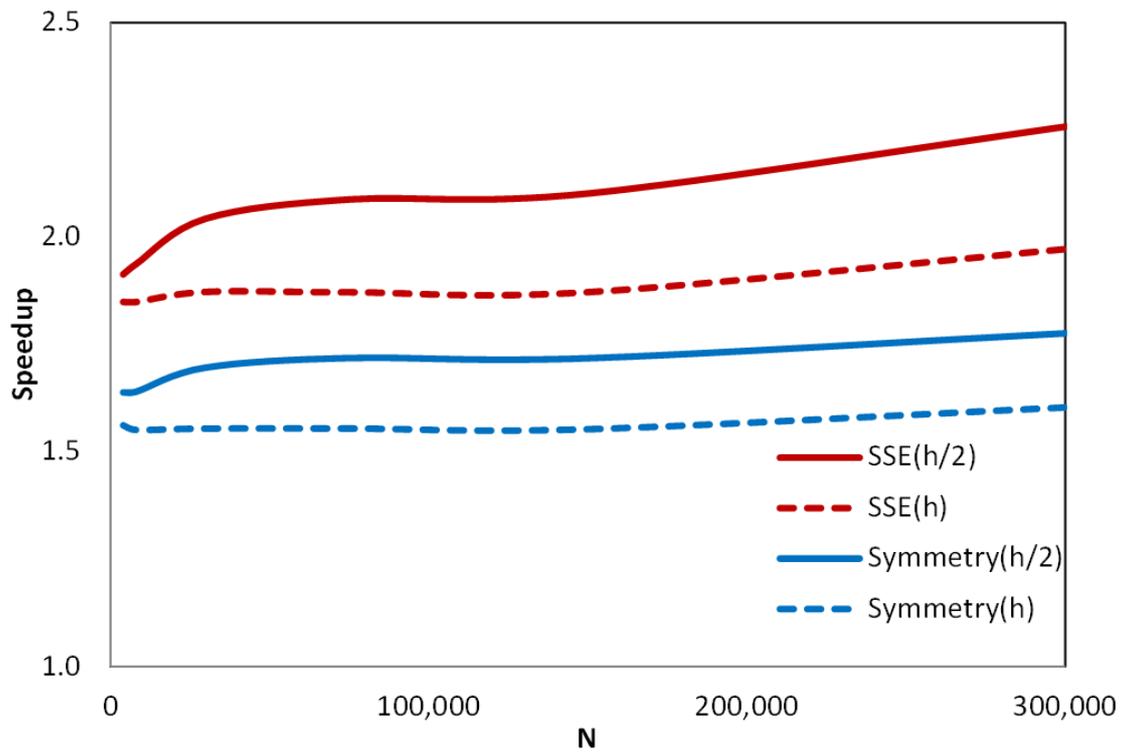

**Fig. 13**. Speedup achieved on CPU for different number of particles (N) when applying symmetry, the use of SSE instructions. Two different cell sizes (*h* and *h/2*) were considered.

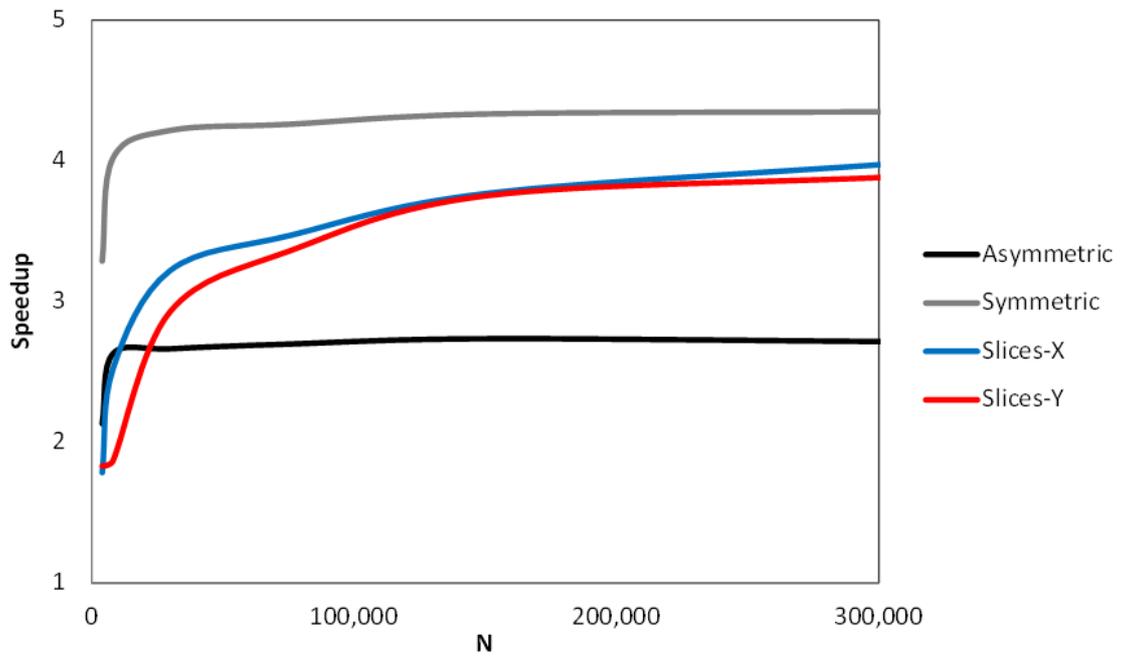

**Fig. 14**. Speedup achieved on CPU for different number of particles (N) with different OpenMP implementations (using 8 logical threads) in comparison with the most efficient single-core version that includes all the previous optimizations.

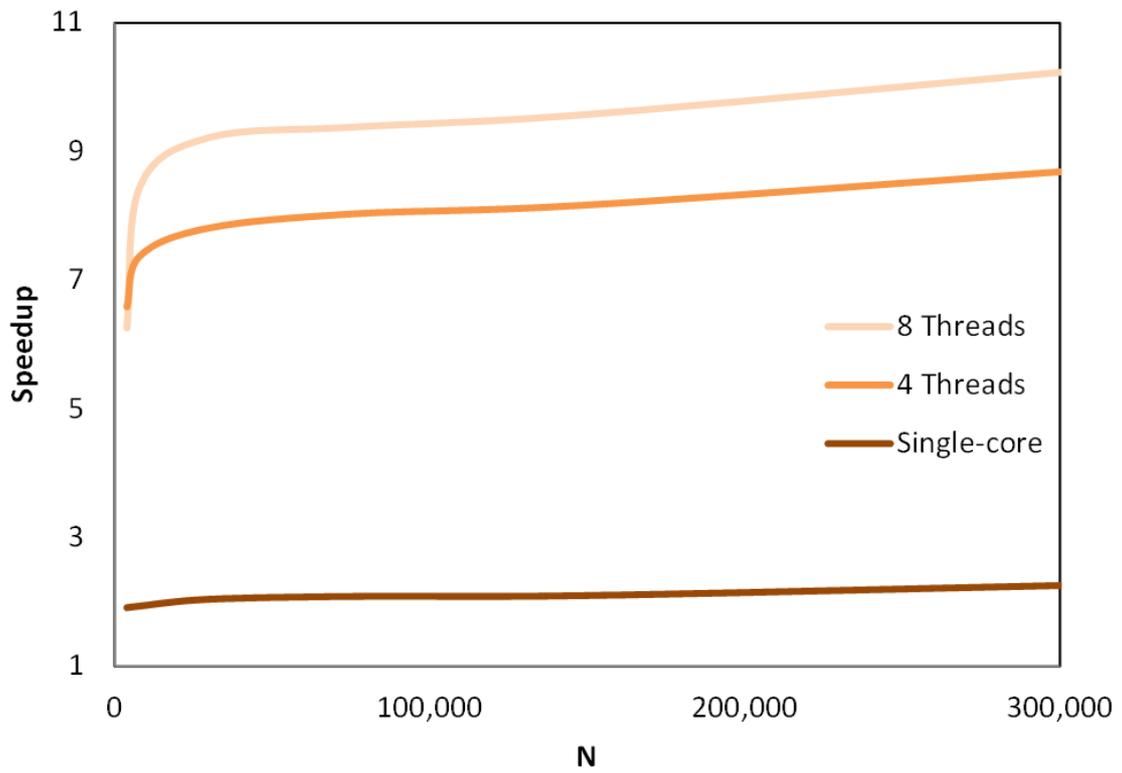

**Fig. 15**. Speedup achieved on CPU for different number of particles (N) with the most efficient implementations (OpenMP and single-core including all the CPU optimizations) in comparison with the single-core implementation without optimizations.

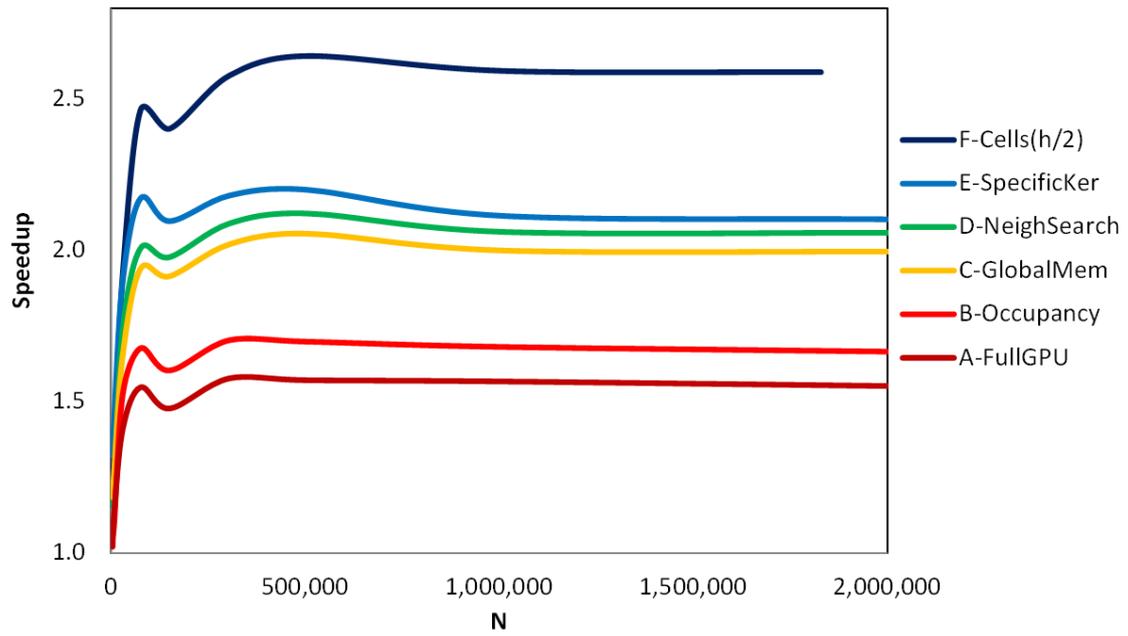

**Fig. 16**. Speedup achieved on GPU for different number of particles (N) when applying the different GPU optimizations using GTX 480. Note that each optimization includes all the previous ones.

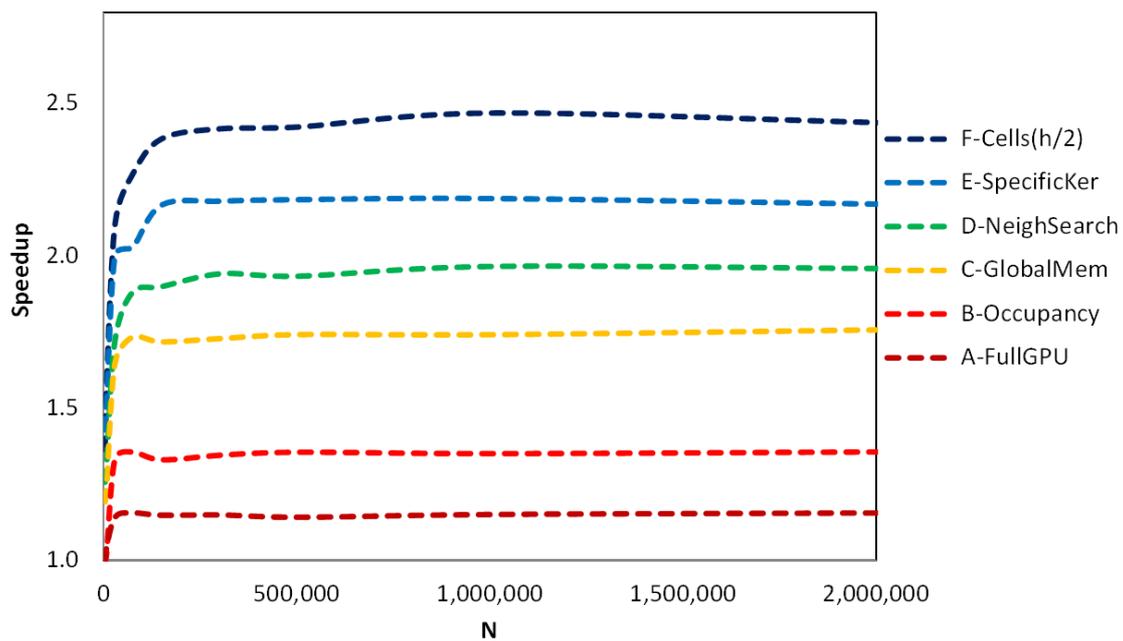

**Fig. 17.** Speedup achieved on GPU for different number of particles (N) when applying the different GPU optimizations using Tesla 1060. Note that each optimization includes all the previous ones.

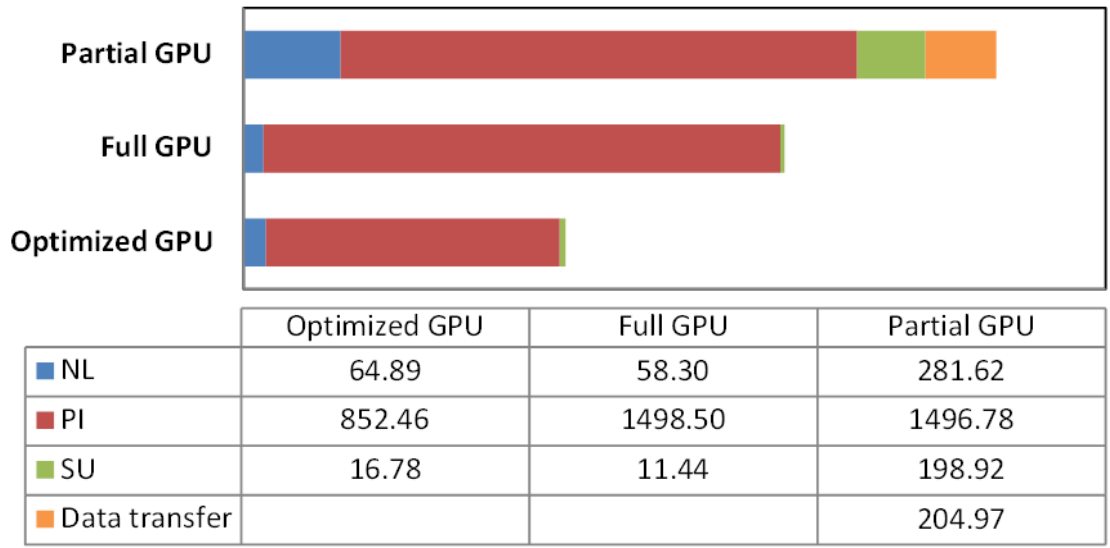

|  | Optimized GPU | Full GPU | Partial GPU |
|---|---|---|---|
| ■ NL | 64.89 | 58.30 | 281.62 |
| ■ PI | 852.46 | 1498.50 | 1496.78 |
| ■ SU | 16.78 | 11.44 | 198.92 |
| ■ Data transfer |  |  | 204.97 |

**Fig. 18**. Computational runtimes (in seconds) using GTX 480 for different GPU optimizations (partial, full and optimized). The testcase described above was simulated with 500,000 particles.

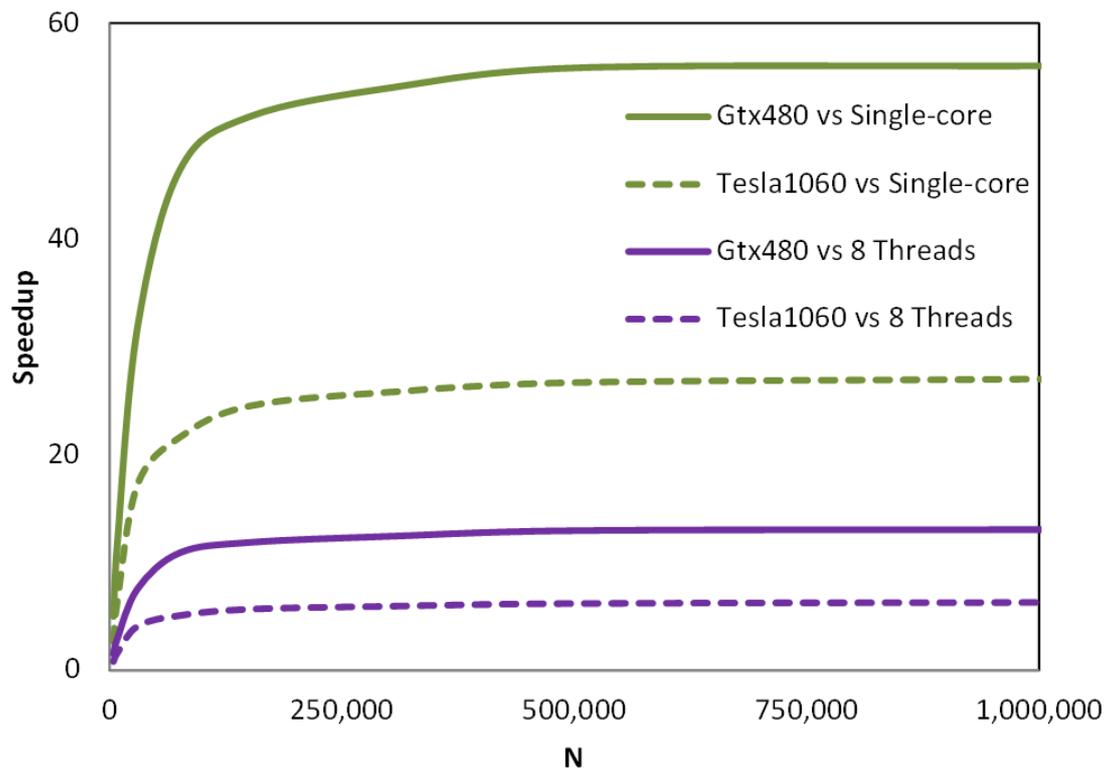

**Fig. 19**. Speedup achieved with the most efficient GPU implementation versus the optimized CPU implementations (single-core and multi-core with 8 threads) using GTX 480 and Tesla 1060 for different number of particles (N).

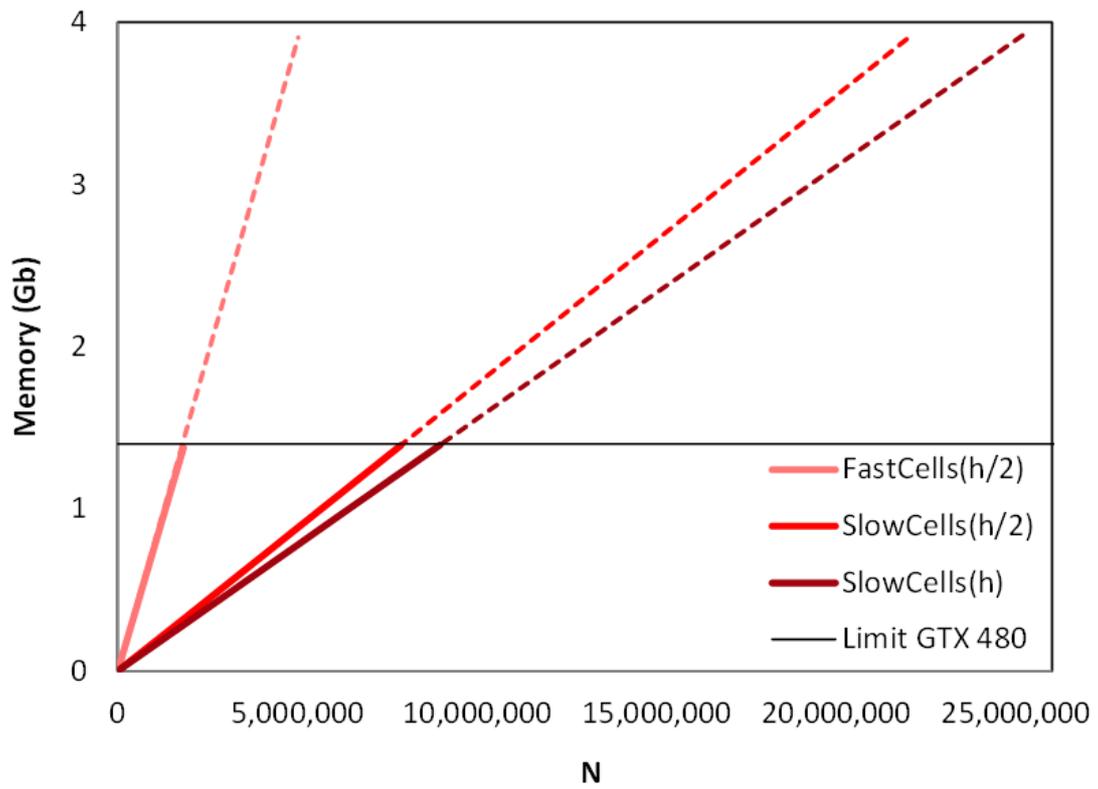

**Fig. 20**. Memory usage for different GPU versions implemented in DualSPHysics.

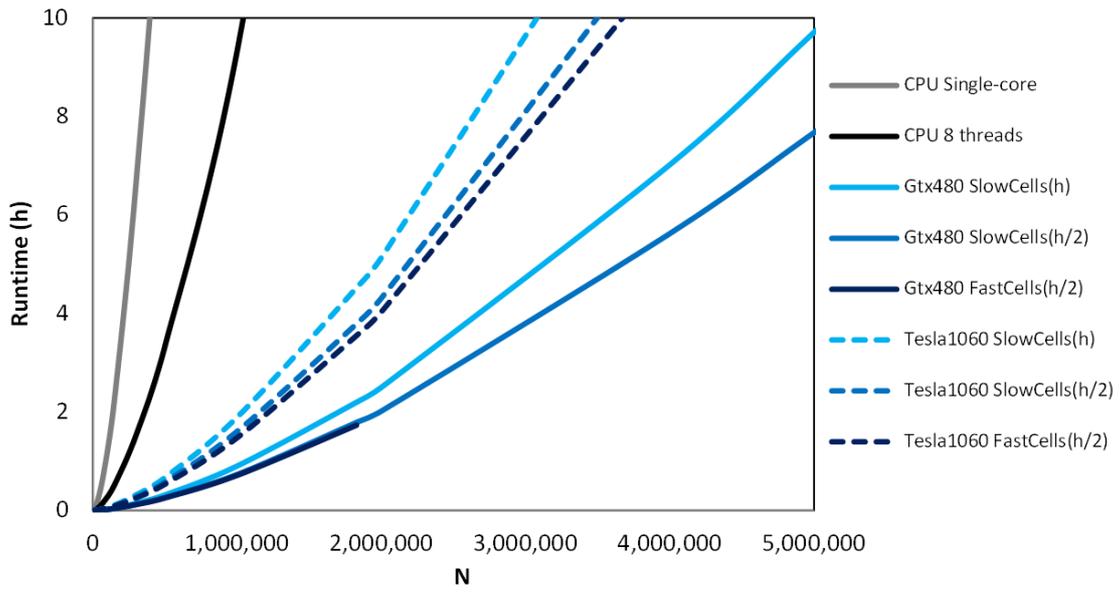

**Fig. 21**. Runtimes for different CPU and GPU implementations.

# Tables

**Table 1.** SPH formulation and references.

| Time integration scheme | Verlet [26] |
|---|---|
| Time Step | Variable [25] |
| Kernel Function | Cubic Spline Kernel [27] |
| Viscosity Treatment | Artificial (alpha=0.25) [28] |
| Equation of State | Tait equation [29] |
| Boundary Condition | Dynamic Boundaries [30] |

**Table 2.** Technical specifications of GPUs according to the compute capability.

| *Technical specifications* | 1.0 | 1.1 | 1.2 | 1.3 | 2.x |
|---|---|---|---|---|---|
| Max. of threads per block | 512 | 512 | 512 | 512 | 1024 |
| Max. of resident blocks per SM | 8 | 8 | 8 | 8 | 8 |
| Max. of resident warps per SM | 24 | 24 | 32 | 32 | 48 |
| Max. of resident threads per SM | 768 | 768 | 1024 | 1024 | 1536 |
| Max. of 32-bit registers per SM | 8 K | 8 K | 16 K | 16 K | 32 K |

**Table 3.** List of variables needed to calculate forces.

| Variable | Size (bytes) | Description |
|---|---|---|
| pos | 3 x 4 | Position in X,Y and Z |
| vel | 3 x 4 | Velocity in X,Y and Z |
| rhop | 4 | Density |
| csound | 4 | Speed of sound |
| prrhop | 4 | Ratio between pressure and density |
| tensil | 4 | Tensile correction following [34] |

**Table 4.** Results of the CPU and GPU simulations.

| Version | Number of particles | Total simulation time | Number of Steps | Computed steps per second | Speedup vs. CPU Single-core | Speedup vs. CPU 8 Threads |
|---|---|---|---|---|---|---|
| CPU Single-core | 503,492 | 14.6 h | 19,855 | 0.4 | *1.0x* | -- |
| | 1,011,354 | 40.7 h | 26,493 | 0.2 | *1.0x* | -- |
| CPU 8 Threads | 503,492 | 3.2 h | 19,806 | 1.7 | 4.6x | *1.0x* |
| | 1,011,354 | 9.1 h | 26,511 | **4.5x** | **4.5x** | *1.0x* |
| GPU Tesla 1060 | 503,492 | 0.5 h | 19,832 | 10.2 | 26.8x | 5.8x |
| | 1,011,354 | 1.5 h | 26,509 | 4.9 | 27.3x | 6.1x |
| GPU GTX 480 | 503,492 | 0.3 h | 19,830 | 21.2 | 55.7x | 12.2x |
| | 1,011,354 | 0.7 h | 26,480 | 10.1 | **56.2x** | **12.5x** |